# Computing Hypergraph Ramsey Numbers by Using Quantum Circuit

Ri Qu, Zong-shang Li, Juan Wang, Yan-ru Bao, Xiao-chun Cao

*School of Computer Science & Technology, Tianjin University, Tianjin 300072, China*

Gaitan and Clark [Phys. Rev. Lett. 108, 010501 (2012)] have recently shown a quantum algorithm for the computation of the Ramsey numbers using adiabatic quantum evolution. We present a quantum algorithm to compute the two-color Ramsey numbers for $r$-uniform hypergraphs by using the quantum counting circuit.

**1 Introduction**

Ramsey numbers are derived form the party problem. In an arbitrary party of $N$ persons, one might ask whether there is a group of $m$ persons who are all mutually acquainted, or a group of $n$ persons who are all mutual strangers. Ramsey theory [1] shows that there exists a threshold value $R(m,n)$ such that for $N \geq R(m,n)$ all parties of $N$ persons will either contain $m$ mutual acquaintances, or $n$ mutual strangers. The threshold value $R(m,n)$ is example of a two-color Ramsey number. One can use an $N$-vertex graph to represent the $N$-person party problem and describe Ramsey numbers by using the language of graph theory [2].

There have already been many types of Ramsey numbers including specific graph, hypergraph, multicolor Ramsey numbers, and so on [3]. We will focus on two-color Ramsey numbers for $r$-uniform hypergraphs. $R(m, n; r)$ is defined as the least integer $N$ such that, in any coloring with two colors (red and blue) of $r$-subsets of the set of $N$ vertices, the set of red $r$-subsets or the set of blue $r$-subsets contain a complete $r$-uniform subhypergraph with $m$ or $n$ vertices. It can also be equivalently shown that for any integer $N$, $N \geq R(m,n;r)$ if and only if every $r$-uniform hypergraph with $N$ vertices contains a complete $r$-uniform subhypergraph with $m$ vertices, or an $n$-independent set. $R(m, n; r)$ is an example of the two-color Ramsey numbers for $r$-uniform hypergraphs. In particular, if $r = 2$ then $R(m,n;r) = R(m,n)$. Ramsey numbers grow extremely quickly such that it is notoriously difficult to calculate them. In fact, for $R(m,n)$ with $m, n \geq 3$ only nine are presently known [3], while for $R(m,n;r)$ with $\min(m,n) > r \geq 3$ only one is known, i.e., $R(4,4;3) = 13$ [4].

Ref. [5] has recently presented a quantum algorithm for the computation of the Ramsey numbers $R(m, n)$ using the adiabatic quantum evolution (AQE) [6]. Ref. [7] shows that $R(m, n; r)$ for $r$-uniform hypergraphs can be computed by using the adiabatic quantum evolution. In this paper, we will present a quantum algorithm to compute $R(m, n; r)$ by using the quantum counting circuit [8, 9].

This paper is organized as follows. In Sec. 2 we will show how the computation of $R(m, n; r)$ can be mapped to a combinatorial optimization problem. In Sec. 3 we first re-describe the above problem by the search problem. Then we use the quantum counting circuit to solve the re-described optimization problem. We summarize the results in Sec. 4.

**2 Optimization problem**

For given integers $N$ and $r$, we can establish a *1-1* corrsepondence between the set of $r$-uniform hypergraphs with $N$ vertices and $\{0,1\}^{B(N,r)}$ where $B$ denotes the binomial coefficient. For convenience, let $V \equiv \{1, 2, ..., N\}$ be the set of $N$ vertices. We define the set of $r$-uniform hypergraphs with $N$ vertices by $P_N^r \equiv \wp(\{e \mid e \subseteq V \wedge |e| = r\})$ where $\wp(A)$ and $|A|$ denote the power set and the cardinality of the set $A$ respectively. For every hypergraph $G \in P_N^r$, there exists a unique $\underbrace{N \times N \times \cdots \times N}_{r \ copies}$ matrix $A(G)$ with the element

$$a_{i_1, i_2, ..., i_r} = \begin{cases} 1 & \{i_1, i_2, ..., i_r\} \in G \\ 0 & \{i_1, i_2, ..., i_r\} \notin G \end{cases}, \quad (1)$$

where $i_1, i_2, ..., i_r \in V$. It is known that $a_{i_1, i_2, ..., i_r} = a_{\delta(i_1, i_2, ..., i_r)}$, where $\delta(i_1, i_2, ..., i_r)$ denotes the permutation of $i_1, i_2, ..., i_r$. Moreover, if $|\{i_1, i_2, ..., i_r\}| < r$ then $a_{i_1, i_2, ..., i_r} = 0$. Thus we can construct a *1-1* correspondence $g_{N,r} : P_N^r \to \{0,1\}^{B(N,r)}$ which satisfies

$$\forall G \in P_N^r, g_{N,r}(G) = a_{i_1^1, i_1^2, ..., i_1^r} a_{i_2^1, i_2^2, ..., i_2^r} \cdots a_{i_{B(N,r)}^1, i_{B(N,r)}^2, ..., i_{B(N,r)}^r}, \quad (2)$$

where $\forall k, j \in \{1, 2, ..., B(N,r)\}$, $i_k^1, i_k^2, ..., i_k^r \in V$, $i_k^1 > i_k^2 > ... > i_k^r$, and there exists $t \in \{1, 2, ..., r\}$ such that $i_j^r = i_k^r, i_j^{r-1} = i_k^{r-1}, ..., i_j^{t+1} = i_k^{t+1}, i_j^t < i_k^t$ if and only if $j < k$. We call $\{i_k^1, i_k^2, ..., i_k^r\}$ by the $k$-th $r$-subset of $V$. By (2), we can get $a_{i_1^1, i_1^2, ..., i_1^r} = a_{r, r-1, ..., 1}$, $a_{i_2^1, i_2^2, ..., i_2^r} = a_{r+1, r-1, ..., 1}$ ,..., $a_{i_{B(N,r)}^1, i_{B(N,r)}^2, ..., i_{B(N,r)}^r} = a_{N, N-1, ..., N-r+1}$ .

If $r=2$, $g_{N,2}(G) = a_{2,1}a_{3,1}\cdots a_{N,1}a_{3,2}a_{4,2}\cdots a_{N,2}\cdots a_{N,N-1}$ which is just the string shown in [5]. For $r=3$, the binary string is

$$g_{N,3}(G) = a_{3,2,1}a_{4,2,1}\cdots a_{N,2,1}a_{4,3,1}a_{5,3,1}\cdots a_{N,3,1}\cdots a_{N,N-1,1} \quad (3)$$
$$a_{4,3,2}a_{5,3,2}\cdots a_{N,3,2}a_{5,4,2}a_{6,4,2}\cdots a_{N,4,2}\cdots a_{N,N-1,2}\cdots a_{N,N-1,N-2}$$

For any $g_{N,r}(G)$, we choose $m$ vertices $S_\alpha = \{k_1, k_2, ..., k_m\}$ from $V$ and construct the product

$$C_\alpha = \prod_{i_1,i_2,...,i_r \in S_\alpha}^{i_1 > i_2 > ... > i_r} a_{i_1,i_2,...,i_r}. \quad (4)$$

Note that $C_\alpha = 1$ if and only if $S_\alpha$ is corresponding to a complete $r$-uniform subhypergraph with $m$ vertices. We can get the sum

$$C_m\left[g_{N,r}(G)\right] = \sum_{\alpha=1}^{B(N,m)} C_\alpha \quad (5)$$

which equals the number of complete $r$-uniform subhypergraphs with $m$ vertices in $G$. Subsequently, choose $n$ vertices $T_\alpha = \{k_1, k_2, ..., k_n\}$ from $V$ and form the product

$$I_\alpha = \prod_{i_1,i_2,...,i_r \in T_\alpha}^{i_1 > i_2 > ... > i_r} \left(1 - a_{i_1,i_2,...,i_r}\right). \quad (6)$$

It is clear that $I_\alpha = 1$ if and only if $T_\alpha$ is corresponding to an $n$-independent set. We can get the num

$$I_n\left[g_{N,r}(G)\right] = \sum_{\alpha=1}^{B(N,n)} I_\alpha \quad (7)$$

which equals the number of $n$-independent sets in $G$. Then we define

$$h_{m,n}\left[g_{N,r}(G)\right] \equiv C_m\left[g_{N,r}(G)\right] + I_n\left[g_{N,r}(G)\right]. \quad (8)$$

It follows that $h_{m,n}\left[g_{N,r}(G)\right] = 0$ if and only if $G$ does not contain a complete $r$-uniform subhypergraph with $m$ vertices or an $n$-independent set.

Just as Ref. [5] and [7], we can use $h_{m,n}\left[g_{N,r}(G)\right]$ as the cost function for the combinatorial optimization problem as follows. For given integers $N$, $m$, $n$ and $r$ we can find a binary string $s \in \{0,1\}^{B(N,r)}$ corresponding to the hypergraph $G_*$ with $N$ vertices that yields the global minimum of $h_{m,n}(x)$ over all $x \in \{0,1\}^{B(N,r)}$. It is clear that $h_{m,n}(s) > 0$ if and only if $N \geq R(m,n;r)$. Thus, for given integers $m$, $n$, and $r$ we begin with $N < R(m,n;r)$ which implies $h_{m,n}(s) = 0$ for $N$, and increment $N$ by 1 until we first find $h_{m,n}(s) > 0$. Then the corresponding $N$ will just equal to $R(m,n;r)$.

## 3 Quantum algorithm

In this section, we will use a quantum circuit to solve the above combinational optimization problem which can be solved by using the AQE algorithm shown in [5, 7].

### 3.1 Re-described problem

For given integers $N$, $m$, $n$ and $r$ we define a Boolean function $f_{m,n}: \{0,1\}^{B(N,r)} \to \{0,1\}$ which satisfies

$$\forall x \in \{0,1\}^{B(N,r)}, f_{m,n}(x) = \begin{cases} 1 & h_{m,n}(x) = 0 \\ 0 & Otherwise \end{cases} \quad (9)$$

Then it is clear that $|f_{m,n}^{-1}(1)| = 0$ if and only if $N \geq R(m,n;r)$. Thus the above combinational optimization problem can be re-described as follows. If we begin with $N < R(m,n;r)$ which implies $|f_{m,n}^{-1}(1)| > 0$ for $N$, and increment $N$ by 1 until we first find $|f_{m,n}^{-1}(1)| = 0$, then the corresponding $N$ will exactly be $R(m,n;r)$. Thus it is important to identify $|f_{m,n}^{-1}(1)|$. We can think of the above Boolean function show in (9) as a particular instance of the search problem of $2^{B(N,r)}$ elements. And $x$ is a solution to the search problem if and only if $f_{m,n}(x) = 1$. The search problem has exactly $|f_{m,n}^{-1}(1)|$ solutions. It is known that the number of solutions to the search problem can be identified by using quantum counting algorithm [8, 9] where a quadratic speedup over the best classical one is shown.

### 3.2 Quantum gates

In this paper, we use some types of quantum gates shown in Fig. 1 to solve the above problem. We define the quantum operations, i.e., $X \equiv \begin{bmatrix} 0 & 1 \\ 1 & 0 \end{bmatrix}$, $H \equiv \frac{1}{\sqrt{2}} \begin{bmatrix} 1 & 1 \\ 1 & -1 \end{bmatrix}$, $Z \equiv \begin{bmatrix} 1 & 0 \\ 0 & -1 \end{bmatrix}$ and $E \equiv \begin{bmatrix} -1 & 0 \\ 0 & -1 \end{bmatrix}$ to respectively represent the quantum NOT, Hadamard, $Z$ and $E$ gates. Suppose $U$ is an $s$-qubit unitary operation. $\wedge^h(U)$ denotes a controlled $U$ gate with $h$ control qubits which is defined by the equation

$$\wedge^h(U)|u_1 u_2 \ldots u_h\rangle|\psi\rangle = |u_1 u_2 \ldots u_h\rangle U^{\prod_{i=1}^{h} u_i}|\psi\rangle \quad (10)$$

where $u_1 u_2 \ldots u_h \in \{0,1\}^h$ and $|\psi\rangle$ is an $s$-qubit state. Thus $\wedge^h(U)$ is an $h+s$-qubit gate, and its first $h$ qubits are called by control qubits while others are target qubits. Similarly, a controlled $U$ gate $\vee^h(U)$ is defined by

$$\vee^h(U)|u_1 u_2 \ldots u_h\rangle|\psi\rangle = |u_1 u_2 \ldots u_h\rangle U^{\prod_{i=1}^{h}(1-u_i)}|\psi\rangle \quad (11)$$

where $u_1 u_2 \ldots u_h \in \{0,1\}^h$ and $|\psi\rangle$ is an $s$-qubit state.

### 3.3 Quantum circuit

We first construct a quantum reversible circuit for the unitary operation $U_{f_{m,n}}$ to transform $|x,y\rangle$ into $|x, y \oplus f_{m,n}(x)\rangle$ as shown in Fig. 2. All qubits in the circuit are classified into three parts: the input qubits, one output qubit and the ancilla qubits. The first $B(N,r)$ qubits are all called by input ones which respectively correspond to bits of $x \equiv x_1 x_2 \cdots x_{B(N,r)} \in \{0,1\}^{B(N,r)}$. Notice that for any $k \in \{1, 2, \ldots, B(N,r)\}$, $|x_k\rangle$ is related with the $k$-th $r$-subset of $V$, and we call the $r$-subset by the related $r$-subset of $|x_k\rangle$. Thus we can identify $x$ with the computational basis state $|x\rangle$. The last qubit is the output one used to store the result of $y \oplus f_{m,n}(x)$. Others are the ancilla qubits used to provide workspace for the computation. The number of ancilla qubits is $B(N,m) + B(N,n) + 3$. And all of ancilla qubits are initialized to $|0\rangle$.

By (10), we can use a controlled NOT gate $\wedge^{B(m,r)}(X)$ to compute $C_\alpha$ in (4). We choose $B(m,r)$ qubits from all input ones as the control qubits of $\wedge^{B(m,r)}(X)$. For any input qubit, if its related $r$-subset is the subset of $S_\alpha$ then it is one of control qubits of $\wedge^{B(m,r)}(X)$. We choose one from the first $B(N,m)$ ancilla qubits as the target qubit of $\wedge^{B(m,r)}(X)$ which is used to store the value of $C_\alpha$. Since the action of $\wedge^{B(m,r)}(X)$ is that the state of control qubits is $\underbrace{|11\ldots 1\rangle}_{B(m,r) \text{ copies}}$ if and only if the state of the target qubit changes from $|0\rangle$ to $|1\rangle$, the output state of the target qubit of $\wedge^{B(m,r)}(X)$ is $|C_\alpha\rangle$. Subsequently, we form a quantum circuit to determine whether $C_m(x)$ in (5) is equal to zero or not. The quantum circuit contains $B(N,m)$ controlled NOT gates $\wedge^{B(m,r)}(X)$ and one $\vee^{B(N,m)}(X)$. All control qubits of $\vee^{B(N,m)}(X)$ consist of the target qubits of all $\wedge^{B(m,r)}(X)$, which means that the first $B(N,m)$ ancilla qubits correspond to the control qubits of $\wedge^{B(N,m)}(X)$. We use the $B(N,m)+1$st ancilla qubit as the target qubit of $\wedge^{B(N,m)}(X)$. By (11), the function of $\wedge^{B(N,m)}(X)$ is that the state of control qubits of $\wedge^{B(N,m)}(X)$ is $\underbrace{|00...0\rangle}_{B(N,m)\ copies}$ if and only if the state of the target qubit changes from $|0\rangle$ to $|1\rangle$. Thus the output state of target qubit of $\wedge^{B(N,m)}(X)$ is $|1\rangle$ if and only if $C_m(x)=0$.

Similarly, we construct a quantum circuit to determine whether $I_n(x)$ in (7) is equal to zero or not by using $B(N,n)$ controlled NOT gates $\wedge^{B(n,r)}(X)$ and one $\wedge^{B(N,n)}(X)$ whose control qubits are respectively corresponding to those from the $B(N,m)+2$nd ancilla quibit to the $B(N,m)+B(N,n)+2$nd one. We use one $\wedge^2(X)$ called by the Toffoli gate to obtain $f_{m,n}(x)$. The control qubits of the Toffoli gate respectively correspond to the target qubits of $\wedge^{B(N,m)}(X)$ and $\wedge^{B(N,n)}(X)$, and its target qubit is the last ancilla one. Thus the action of $\wedge^2(X)$ is to change the state of its target qubit from $|0\rangle$ to $|f_{m,n}(x)\rangle$. Then we use one $\wedge^1(X)$ called by the standard controlled NOT to compute $y \oplus f_{m,n}(x)$. The control and target qubits of $\wedge^1(X)$ are respectively the last ancilla and output ones. Finally, we apply the reverse of the circuit used to compute $f_{m,n}(x)$, so that the circuit for $U_{f_{m,n}}$ is the reversible circuit where all ancilla qubits come back to $|0\rangle$.

We can get

$$|x\rangle\frac{|0\rangle-|1\rangle}{\sqrt{2}} \xrightarrow{U_{f_{m,n}}} (-1)^{f_{m,n}(x)}|x\rangle\frac{|0\rangle-|1\rangle}{\sqrt{2}} \qquad (12)$$

Notice that the state of the output qubit is not changed. Thus the action of the operation $U_{f_{m,n}}$ may be written

$$|x\rangle \xrightarrow{U_{f_{m,n}}} (-1)^{f_{m,n}(x)}|x\rangle \qquad (13)$$

We say that $U_{f_{m,n}}$ is the same as the Oracle operation in Grover's search algorithm [8, 10]. Thus we can construct a quantum circuit of the Grover iteration $G_{m,n} = (2|\psi\rangle\langle\psi| - I)U_{f_{m,n}}$ where $|\psi\rangle = \frac{1}{\sqrt{2^{B(N,r)}}} \sum_{i=0}^{2^{B(N,r)}-1} |i\rangle$ and $I = \sum_{i=0}^{2^{B(N,r)}-1} |i\rangle\langle i|$, as shown in Fig. 3.

Quantum counting is an application of the quantum phase estimation algorithm [8, 11] to estimate the eigenvalues of the Grover iteration $G_{m,n}$. The phase estimation circuit used for quantum counting is shown in Fig. 4. The function of the circuit is to estimate $\theta$ with $\sin^2 \theta/2 = |f_{m,n}^{-1}(1)|/2^{B(N,r)}$ to $w$ bits of accuracy, with a probability of success at least $1-\varepsilon$. The first register contains $t \equiv w + \lceil \log(2 + 1/2\varepsilon) \rceil$ qubits, and the second register contains $B(N,r) + B(N,m) + B(N,n) + 4$ qubits which appear in the circuit for $U_{f_{m,n}}$. It is known that an error $\Delta |f_{m,n}^{-1}(1)|$ of the estimate of $|f_{m,n}^{-1}(1)|$ satisfies

$$\Delta |f_{m,n}^{-1}(1)| < 2^{-w} \cdot \left( \sqrt{|f_{m,n}^{-1}(1)| \cdot 2^{B(N,r)}} + 2^{B(N,r)-(w+2)} \right) \tag{14}$$

In this paper, we choose $w = \lceil B(N,r)/2 \rceil + 1$ and $\varepsilon = 1/6$ (Note that we can choose $\varepsilon \to 0$). Then we have $t = \lceil B(N,r)/2 \rceil + 4$, so the algorithm requires $\Theta(\sqrt{2^{B(N,r)}})$ Grover iterations. By (14), the error $\Delta |f_{m,n}^{-1}(1)| < \sqrt{|f_{m,n}^{-1}(1)|}/2 + 1/16$. If $|f_{m,n}^{-1}(1)| = 0$ then we have $\Delta |f_{m,n}^{-1}(1)| < 1/16$, so the algorithm must produce the estimate zero with probability at least $5/6$. Conversely, if $|f_{m,n}^{-1}(1)| > 0$ then it is easy to verify that the estimate for $|f_{m,n}^{-1}(1)|$ is not equal to 0 with probability at least $5/6$. Thus the quantum algorithm for $R(m, n; r)$ by using quantum counting circuit is shown as follows.

*Algorithm 1*. Quantum algorithm for $R(m, n; r)$

Inputs: (i) given integers $m$, $n$ and $r$. (ii) a strict lower bound $L$ for $R(m, n; r)$ which can be found in Ref. [3].

Outputs: $R(m, n; r)$.

Procedure:
(i) $L \to N$.
(ii) For $N$, $m$, $n$ and $r$, construct and perform the above quantum counting circuit to determine whether $|f_{m,n}^{-1}(1)| = 0$ or $|f_{m,n}^{-1}(1)| > 0$ with probability at least $5/6$.

(iii) If $\left|f_{m,n}^{-1}(1)\right| > 0$ which implies $N < R(m,n;r)$, then $N$ is incremented by 1 and go to the step (ii). Otherwise, continue the next step.

(iv) $N \to R(m,n;r)$.

## 4 Conclusions

In this paper, we show how the computation of $R(m, n; r)$ can be mapped to a combinatorial optimization problem. Then we re-describe the optimization problem by the search problem. We present a quantum algorithm to solve the re-described problem by using the quantum counting circuit. Since quantum counting algorithm has a quadratic speedup over the best classical one, our algorithm for $R(m, n; r)$ has the same speedup. For $N$, $m$, $n$ and $r$ the circuit contains $L(N) = \lceil 3 \cdot B(N,r)/2 \rceil + B(N,m) + B(N,n) + 8$ qubits, while the AQE algorithm only requires $P(N) = B(N,r)$ qubits [5, 7]. If $N = R(m,n;r)$, $L[R(m,n;r)]$ is the maximum number of qubits required by the quantum counting. However, the AQE algorithm [5,7] for $R(m, n; r)$ only requires at most $P[R(m, n; r)]$ qubits. Thus our algorithm uses greatly more qubits than the AQE algorithm. For example, if $\min(m,n) > r \geq 3$, the smallest Ramsey number for $r$-uniform hypergraphs is $R(4,4;3) = 13$. Our algorithm for $R(4,4;3) = 13$ requires 1867 qubits while the AQE algorithm needs 286 qubits.

**Ackownlegements**

This work was financially supported by the National Natural Science Foundation of China under Grant No. 61170178.

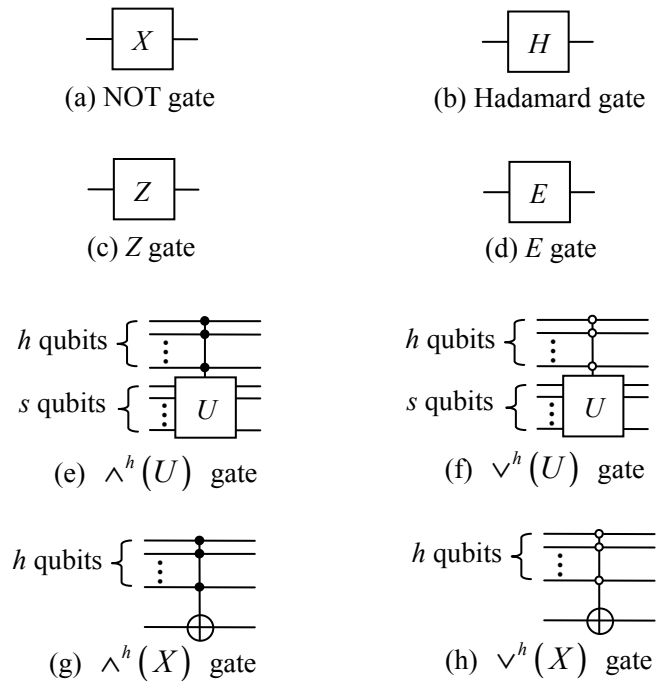

Figure 1. Some types of quantum gates.

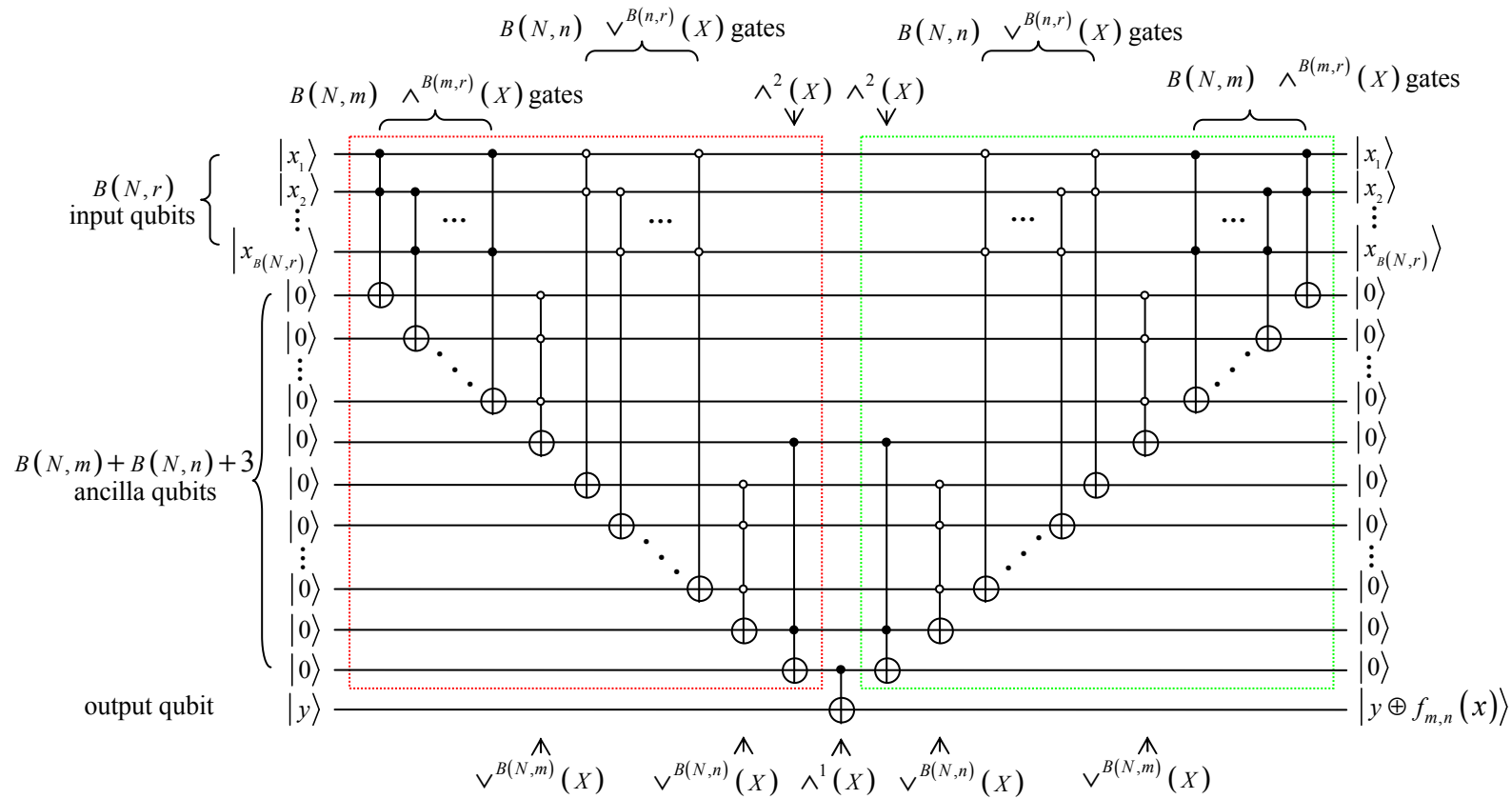

Figure 2. The quantum circuit of the unitary operation $U_{f_{m,n}}$.

The circuit in the red dotted box is used to compute $f_{m,n}$, while the circuit in the green dotted box is the reverse of one in the red dotted box.

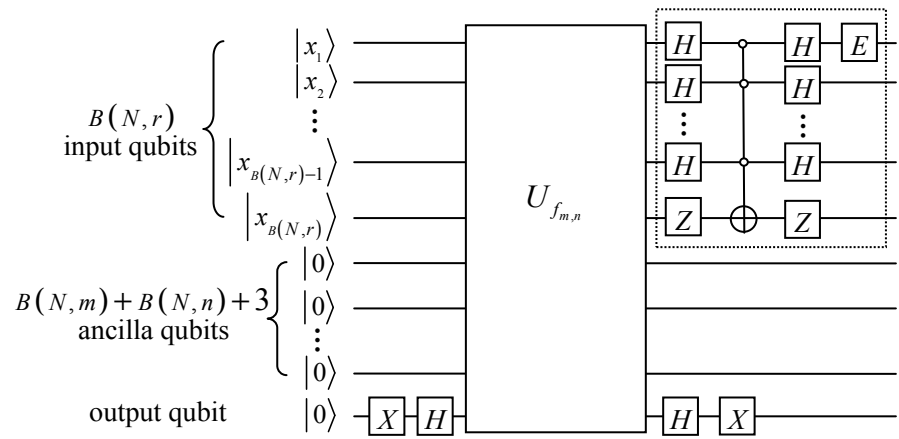

Figure 3. The circuit of the operation $G_{m,n}$. Notice that the circuit in the dotted box is used to perform the operation $2|\psi\rangle\langle\psi| - I$.

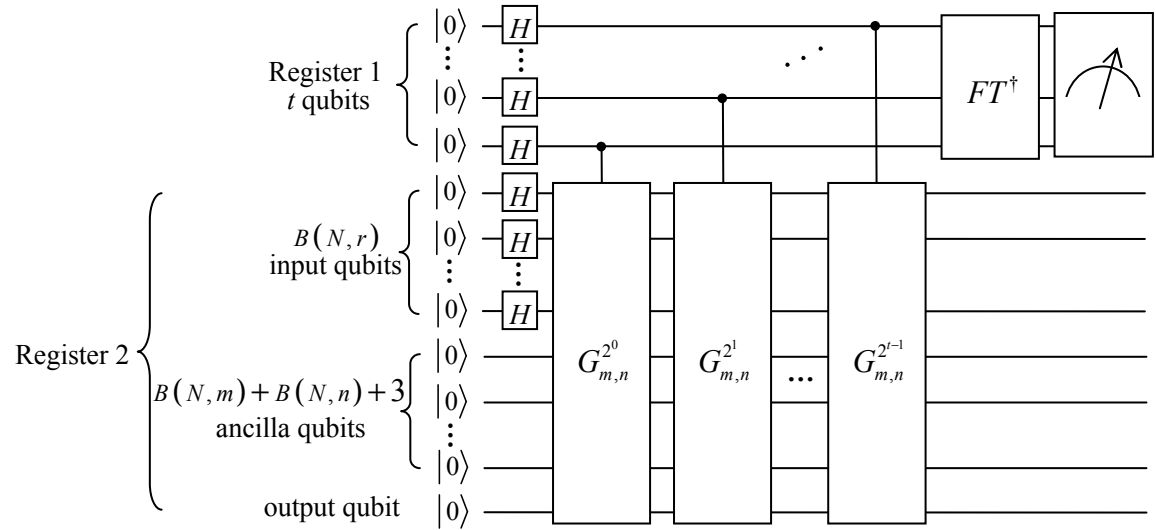

Figure 4. The circuit of the quantum counting. Notice that $FT^\dagger$ is the circuit of the inverse quantum Fourier transform [8] on the first register.